\documentclass[twocolumn]{aastex63}

\received{August 28, 2020}
\revised{September 18, 2020}
\accepted{\today}

\shorttitle{IFS Discovery of a Fast Shell in HuBi\,1}
\shortauthors{Rechy-Garc\'\i a et al.}
\graphicspath{{./}{figures/}}

\begin{document}

\title{Discovery of a Fast Expanding Shell in the Inside-out Born-Again Planetary Nebula HuBi\,1 Through High-Dispersion Integral Field Spectroscopy}

\correspondingauthor{J.\,S.\,Rechy-Garc\'{i}a}
\email{j.rechy@irya.unam.mx}

\author[0000-0002-0121-2537]{J.S.\ Rechy-Garc\'\i a}
\affil{Instituto de Radioastronom\'ia y Astrof\'isica (IRyA),  UNAM 
Campus Morelia, Apartado postal 3-72, 58090, Morelia, Michoacán, Mexico}

\author[0000-0002-7759-106X]{M.A.\ Guerrero}
\affil{
Instituto de Astrof\'\i sica de Andaluc\'\i a, IAA-CSIC, 
Glorieta de la Astronom\'\i a s/n, E-18008, Granada, Spain
}

\author[0000-0003-4946-0414]{E.\ Santamar\'\i a}
\affil{
Instituto de Astronom\'\i a y Meteorolog\'\i a, CUCEI, 
Univ.\ de Guadalajara, 
Av.\ Vallarta 2602, Arcos Vallarta, 44130 Guadalajara, Mexico
}

\author[0000-0001-8252-6548]{V.M.A.\ G\'omez-Gonz\'alez}
\affil{Instituto de Radioastronom\'ia y Astrof\'isica (IRyA),  UNAM 
Campus Morelia, Apartado postal 3-72, 58090, Morelia, Michoacán, Mexico}

\author[0000-0003-2653-4417]{G.\ Ramos-Larios}
\affil{
Instituto de Astronom\'\i a y Meteorolog\'\i a, CUCEI, 
Univ.\ de Guadalajara, 
Av.\ Vallarta 2602, Arcos Vallarta, 44130 Guadalajara, Mexico
}

\author[0000-0002-5406-0813]{J.A.\ Toal\'{a}}
\affil{Instituto de Radioastronom\'ia y Astrof\'isica (IRyA),  UNAM 
Campus Morelia, Apartado postal 3-72, 58090, Morelia, Michoacán, Mexico}

\author[0000-0002-7705-2525]{S.\ Cazzoli}
\affil{
Instituto de Astrof\'\i sica de Andaluc\'\i a, IAA-CSIC, 
Glorieta de la Astronom\'\i a s/n, E-18008, Granada, Spain
}

\author[0000-0003-0242-0044]{L.\ Sabin}
\affil{
Instituto de Astronom\'{i}a, UNAM, Apdo. Postal 877, Ensenada 22860, B.C., Mexico}

\author[0000-0003-0939-8724]{L.F.\ Miranda}
\affil{
Instituto de Astrof\'\i sica de Andaluc\'\i a, IAA-CSIC, 
Glorieta de la Astronom\'\i a s/n, E-18008, Granada, Spain
}

\author[0000-0002-3981-7355]{X.\ Fang}
\affil{Key Laboratory of Optical Astronomy, National Astronomical Observatories, Chinese Academy of Sciences (NAOC), Beijing, China
}
\affil{Department of Physics \& Laboratory for Space Research, Faculty of Science, University of Hong Kong, Hong Kong, China
}

\author{J.\ Liu}
\affil{Key Laboratory of Optical Astronomy, National Astronomical Observatories, Chinese Academy of Sciences (NAOC), Beijing, China
}
\affil{School of Astronomy and Space Sciences, University of Chinese Academy of Sciences (UCAS), Beijing, China
}
\affil{WHU-NAOC Joint Center for Astronomy, Wuhan University, Wuhan, China
}




\begin{abstract}

HuBi\,1 has been proposed to be member of the rare class of {\it born-again} planetary nebulae (PNe), i.e., its central star experienced a {\it very late thermal pulse} and ejected highly-processed material at high speeds inside the old hydrogen-rich PN. 
In this letter we present  
GTC MEGARA integral field spectroscopic observations  of the innermost regions of HuBi\,1 at high spectral resolution $\simeq16$ km~s$^{-1}$
and multi-epoch sub-arcsec images obtained $\simeq 12$ yr apart.  
The analysis of these data indicates that the inner regions of HuBi\,1 were ejected $\simeq200$ yr ago and expand at velocities $\simeq300$ km~s$^{-1}$, in excellent agreement with the {\it born-again} scenario.  
The unprecedented tomographic capabilities of the GTC MEGARA high-dispersion observations used here reveal that the ejecta in HuBi\,1 has a shell-like structure, in contrast to the disrupted disk and jet morphology of the ejecta in other {\it born-again} PNe.

\end{abstract}



\section{Introduction}
\label{sec:intro}

\begin{figure}
\begin{center}
\includegraphics[width=0.95\linewidth]{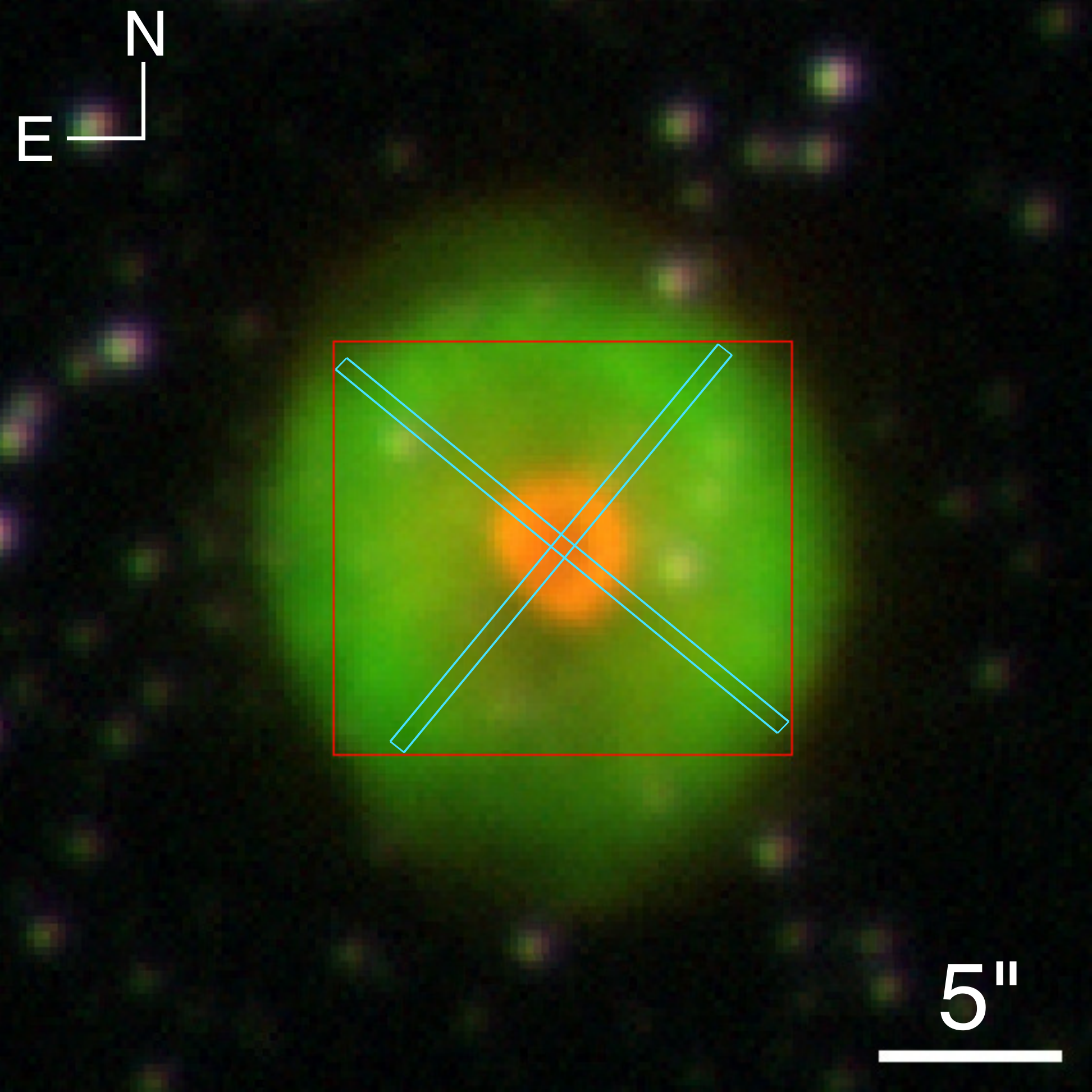} 
\caption{
NOT ALFOSC color-composite image of HuBi\,1 in H$\alpha$ (green), 
[N~{\sc ii}] (red), and $r'$-SDSS (blue).  
The area covered by the GTC MEGARA observations is marked by a red rectangle 
and the location of 
the MEGARA pseudo-slits used in Figure~\ref{fig.PV} in cyan. 
}
\vspace*{-0.45cm}
\label{fig:img}
\end{center}
\end{figure}

The planetary nebula (PN) HuBi\,1 \citep[PN\,G012.2+04.9, a.k.a.\ PM\,1-188;][]{HuBibo1990} 
has recently merited attention because the continuous decline by $\sim10$~mag of its central star (CSPN) in the last 48~yr and its unusual ionization structure \citep{Guerrero2018}.  
Its outer shell has higher ionization than the innermost regions 
(Figure~\ref{fig:img}), opposite to that observed in typical photoionized nebulae, and the latter presents an inverted ionization structure, with emission from low-ionization species of N$^+$, O$^+$, and S$^+$ closer to the central star (CSPN) than that from high-ionization species such as O$^{++}$ and He$^{++}$.  
HuBi\,1 is thus {\it inside-out}.

Using state-of-the-art stellar evolution models \citep{MM2016}, \citet{Guerrero2018} proposed that its CSPN,  a C-rich [Wolf-Rayet] star of spectral type [WC10] \citep{Pena2005}, had experienced a {\it very late thermal pulse} \citep[VLTP;][]{Iben1983}.  
This would make it a member of the exclusive {\it born-again} class and would support a VLTP evolutionary path for [WC] CSPNe.
A VLTP event can occur to CSPNe evolving on the white dwarf cooling track, when H-burning on its surface makes the He shell to reach the critical mass to ignite it into C and O in a short-lived thermonuclear runaway \citep{Herwig2005,Lawlor2006,MillerAlthaus2006}.  
As H-deficient and C-rich material expands, it cools down on short-time scales and forms dust that enshrouds the CSPN 
\citep{Borkowski1994,Evans2006}.

A common feature of the {\it born-again} PNe with detailed spatio-kinematic studies (i.e., A\,30, A\,58, A\,78, and Sakurai's object) is the presence of H-deficient fast bipolar outflows associated with equatorial disk-like ejecta \citep[][]{Pollacco1992,GM1996,Meaburn1996,Chu1997,Meaburn1998,Hinkle2014}. 
The detection of fast moving material in the innermost regions of HuBi\,1 would lend strong support to its {\it born-again} nature, providing a textbook case to investigate the origin of the high turbulence \citep{Acker2002}, enriched C and N abundances \citep{GarciaRojas2013}, and mixed CO chemistry of PNe with [WC] CSPNe \citep{Perea2009}.


The inverted ionization structure of this region provides indirect evidence of shock-excitation caused by a fast outflow interacting with the old PN. 
Moreover, the inspection of the high-dispersion echelle spectra presented by \citet{Guerrero2018} suggests the presence of fast moving material.  
To confirm it, we have obtained multi-epoch narrow-band images 
and high-dispersion integral field spectroscopic (IFS) observations of the inner shell of HuBi\,1.  
These new observations indeed provide evidence of a fast outflow and allow us to determine the spatio-kinematic structure of the innermost regions of HuBi\,1 and to derive its age.  
The observations are described in \S2 and their analyses in \S3.  
A discussion is presented in \S4 and a summary in \S5.



\section{OBSERVATIONS} 
\label{sec:observations}

\begin{figure*}
\begin{center}
\includegraphics[width=\linewidth]{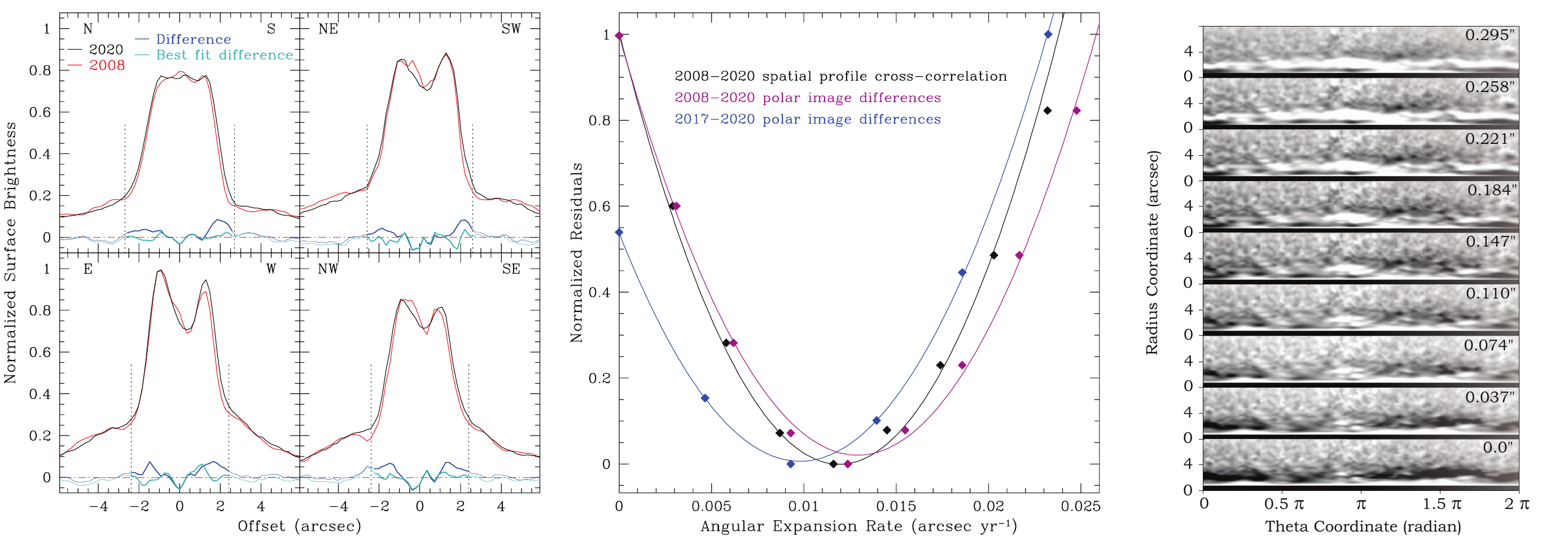}
\caption{
Investigation of the angular expansion of HuBi\,1.  
{\it (left)} 
Stellar-continuum subtracted [N~{\sc ii}] spatial profiles of the inner region of HuBi\,1 along PAs 0$^{\circ}$ (top-left), 45$^{\circ}$ (top-right), 90$^{\circ}$ (bottom-left), and 135$^{\circ}$ (bottom-right) extracted from the 2008 (red line) and 2020 (black line) NOT images.  
Their differences are shown by a dark blue line and those of the best-fit expansion by a cyan line.  
The location of the inner region is marked by vertical black dotted lines and the zero level by a horizontal red dashed line.  
{\it (middle)} 
Normalized residuals between the spatial profiles and polar images of the inner region of HuBi\,1 (diamonds) and least-square best fits (solid lines).
{\it (right)} 
Stack of the [N~{\sc ii}] 2008 and 2020 polar image differences of the inner region of HuBi\,1.  
The angular offset applied to the 2020 polar images is overlaid on the corresponding polar image difference.  
}
\vspace*{-0.25cm}
\label{ang.exp}
\end{center}
\end{figure*}

\subsection{Multi-epoch Optical Imaging}

Multi-epoch images of HuBi\,1 in the [N~{\sc ii}] $\lambda$6584 emission line were obtained at the 2.5~m Nordic Optical Telescope (NOT) at the Observatorio del Roque de los Muchachos (ORM) in La Palma, Spain, using the ALhambra Faint 
Object Spectrograph and Camera (ALFOSC) and the Observatorio de Sierra Nevada (OSN) [N~{\sc ii}] narrow-band filter E16 ($\lambda_c=6583$ \AA, $\Delta\lambda=13$ \AA).
Two 600 s exposures were obtained on 2008 September 2 
using the EEV 2K$\times$2K CCD camera with plate scale of 0\farcs184~pix$^{-1}$ \citep{Guerrero2018}, and three 900 s exposures on 2017 May 27 
and 2020 July 27 
using the E2V 231-42 2K$\times$2K CCD with plate scale of 0\farcs211 pix$^{-1}$.  
In the latest run, three 900 s exposures were obtained through the 
OSN H$\alpha$ filter H01 ($\lambda_c=6565$ \AA, 
$\Delta\lambda=13$ \AA) and five 30 s exposures in $r'$-SDSS. 
In all cases, a dithering of a few arcsec was applied between individual exposures to improve the quality of the final image.

The individual exposures were bias subtracted and flat-fielded using twilight sky frames, and then aligned and combined to remove cosmic rays using standard IRAF routines. 
The spatial resolution of the images, as derived from stars in the field of view (FoV), was 0\farcs65 in 2008 and 0\farcs75 in 2017 and 2020.
A color-composite image of HuBi\,1 using the images obtained in 2020 is presented in 
Figure~\ref{fig:img}.

%
%


\subsection{Integral Field Spectroscopy}

IFS observations were obtained on 2020 August 6 with the Multi-Espectr\'ografo en GTC de Alta Resoluci\'on para Astronom\'ia \citep[MEGARA;][]{GildePaz2018} at the Gran Telescopio de Canarias (GTC) of the ORM.
The integral-field unit (IFU) mode covering a FoV of 12\farcs5$\times$11\farcs3 with 567 hexagonal spaxels of maximal diameter 0\farcs62 was used.  
The High-resolution Volume-Phased Holographic (VPH) grism VPH665-HR used during the observations covers the 6405.6--6797.1 \AA\ wavelength range and provides a spectral dispersion of 0.098 \AA~pix$^{-1}$ and a resolution $R$ of 18,700 (i.e., $\simeq16$ km~s$^{-1}$).  
Three 900 s exposures were obtained to facilitate cosmic rays removal.

The MEGARA raw data were reduced following the Data Reduction Cookbook \citep[Universidad Complutense de Madrid,][]{Pascual2019} using the {\it megaradrp} v0.10.1 pipeline released on 2019 June 29. 
This pipeline applies sky and bias subtraction, flat-field 
correction, wavelength calibration, and spectra tracing and 
extraction. 
The sky subtraction is done using 56 ancillary fibres located $\approx2\farcm0$ from the center of the IFU.
The regularization grid task {\it megararss2cube}\footnote{
Task developed by J.\,Zaragoza-Cardiel available at \url{https://github.com/javierzaragoza/megararss2cube}
} 
was used to produce a final 52$\times$58$\times$4300 data cube with 0\farcs2 square spaxels. 
The spatial resolution of the data after processing was $\simeq1\farcs0$ as derived from the FWHM of stars in the FoV. 
The flux calibration was performed using observations of the spectro-photometric standard HR\,7596 obtained immediately after those of HuBi\,1.

\section{Results} 
\label{sec:results}

The multi-epoch images and high-dispersion IFS data of HuBi\,1 provide us the means to investigate the spatio-kinematics of its innermost regions on the plane of the sky and along the line of sight.

\subsection{Angular Expansion}

An inspection of the 2008 and 2020 [N~{\sc ii}] images of HuBi\,1 is suggestive of the angular expansion of its innermost structure.  
The profile cross-correlation (PCC) and quantified magnification (QM) methods of minimization of residuals between images of different epochs described by  \citet{Santamaria2019} provides a reliable method to determine quantitatively its angular expansion rate \citep{Guerrero2020}. 
In preparation of these analyses, the [N~{\sc ii}] images of the three different epochs were carefully aligned and the pixel scale of the 2017 and 2020 images matched to that of the 2008 image using a sample of well-detected field stars distributed uniformly around HuBi\,1.  
Point spread functions (PSF) of these stars were then compared and the images smoothed using a 2D Gaussian filter to make the PSF of all images similar with a final FWHM of 0\farcs8.

For the PCC method, spatial profiles along four different directions (PA=0$^\circ$, 45$^\circ$, 90$^\circ$, and 135$^\circ$) were extracted from the 2008 and 2020 images using a 2 pixel-wide ($\simeq0\farcs37$) rectangular aperture. 
The emission from the CSPN in the 2008 image, not present in the 2020 image, was modeled and removed using the averaged PSF of stars in the FoV.  
The comparison of the 2020 (black line) and 2008 (red line) spatial profiles in Figure~\ref{ang.exp}-left confirms the angular expansion, with a noticeable excess of emission in the difference profile (blue line). 
The spatial profile of the 2008 image was progressively shifted outwards to simulate its angular expansion and subtracted from that of 2020.
The residuals of the differences of these two profiles at the location of the inner structure decrease until they reach a minimum value and then increase again.  
The normalized residuals are plotted in Figure~\ref{ang.exp}-middle together with a quadratic least-square fit.   
The best-fit implies an expansion rate 0\farcs0119~yr$^{-1}$, which results in the residuals shown as a cyan line in Figure~\ref{ang.exp}-left.

For the QM method, we note that the expansion rate derived above implies magnification factors of a few percent that would artificially broaden the inner shell of HuBi\,1.  
To avoid those effects,  the 2008 and 2020 images were transformed into polar coordinates and the 2020  image shifted along the radial coordinate to simulate an angular expansion and subtracted from the 2008 image (Fig.~\ref{ang.exp}-right).  
The values of the dispersion in the difference maps in a box with radius $1\farcs0 \leq r \leq 3\farcs0$ that excises the emission from the CSPN are normalized and plotted in the middle panel of Figure~\ref{ang.exp}.  
As in the previous case, the dispersion decreases until it reaches a minimum value and then increases again, with a quadratic least-square fit implying an expansion rate 0\farcs0129~yr$^{-1}$.  
This corresponds to a shift $\simeq0\farcs153$ between the 2008 and 2020 images, close to the central polar image difference in Figure~\ref{ang.exp}-right.

The QM method was also applied to the pair of [N~{\sc ii}] images obtained in 2017 and 2020.  
These are closer in time, but they are not affected by the emission from the CSPN.  
In this case, the quadratic least-square fit to the dispersion corresponds to an expansion rate 0\farcs0098~yr$^{-1}$ (Fig.~\ref{ang.exp}-middle).

The three estimates of the angular expansion rate of the inner shell of 
HuBi\,1 derived above are rather consistent (Fig.~\ref{ang.exp}-middle).  
Hereafter we will adopt an averaged angular expansion rate of 0\farcs0115$\pm$0\farcs0016~yr$^{-1}$, implying a velocity on the plane of the sky (55$\pm$8)$\times d$ km~s$^{-1}$, where $d$ is the distance in kpc.
Adopting a distance of 5.3$\pm$1.3 kpc derived using the method described by \citet{Frew2016} with improved values for the nebular radius and H$\alpha$ flux \citep{Guerrero2018}, 
this expansion velocity is 290$\pm$80 km~s$^{-1}$. 

\subsection{Kinematics}

\begin{figure}
\begin{center}
\includegraphics[bb=30 155 570 580,width=0.95\linewidth]{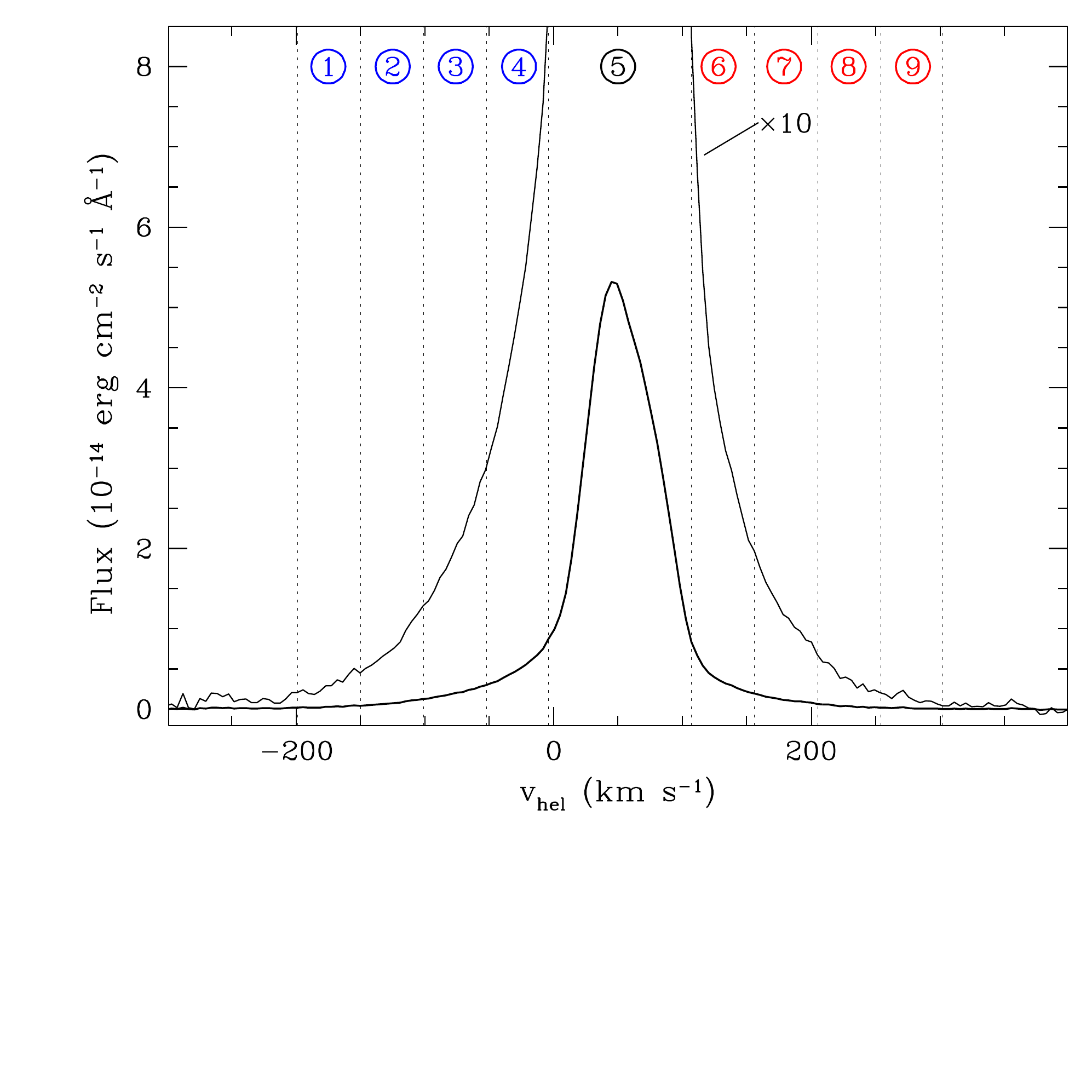}
\caption{
GTC MEGARA emission profile of the [N~{\sc ii}] $\lambda$6584 line of the inner shell of HuBi\,1 integrated within a circular region 2\farcs5 in radius.  
The emission profile is multiplied by 10 to show more clearly the broad emission line wings.  
The vertical dotted lines mark the velocity range of the nine panels in 
Figure~\ref{fig.MEGARA} as also labeled here.  
}
\vspace*{-0.45cm}
\label{fig:spec1D}
\end{center}
\end{figure}

\begin{figure*}
\begin{center}
\vspace*{-0.2cm}
\includegraphics[width=0.95\linewidth]{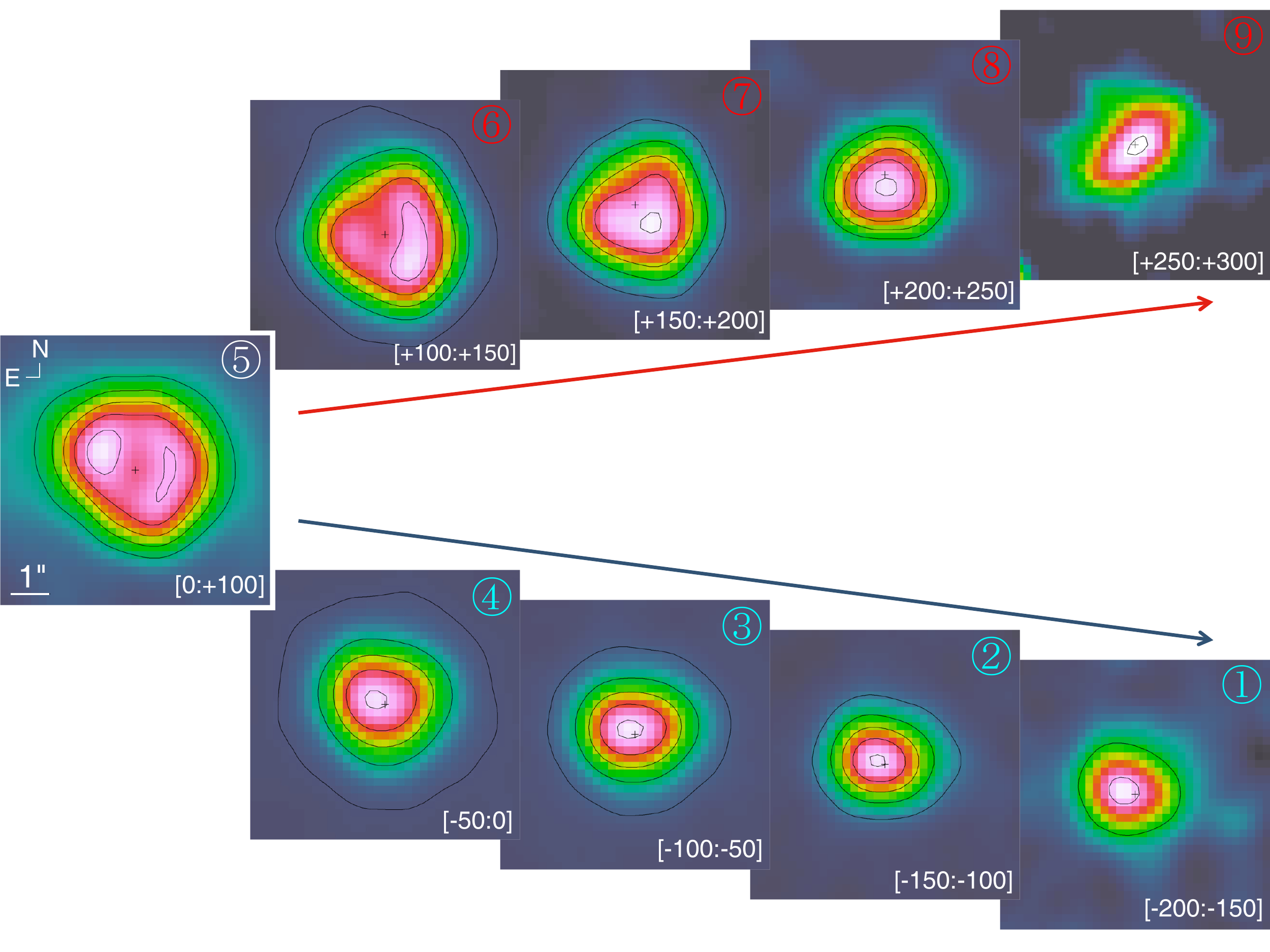}
\vspace*{-0.4cm}
\caption{
GTC MEGARA tomography of the inner shell of HuBi\,1 in the [N~{\sc ii}] $\lambda$6584 emission line.  
The heliocentric velocity range labeled at the bottom-right of each panel corresponds to the velocity ranges defined in Figure~\ref{fig:spec1D} according to the top-right labels.
The leftmost panel corresponds to the systemic velocity.
The arrows indicate increasing velocity difference with respect to the systemic velocity.
The ``cross'' marks the location of the center of the inner shell.  
In all channels, except the one at the systemic velocity, which is contaminated by the emission from the outer shell, the surface brightness of the most extended contour is  3$\times$10$^{-17}$ erg~cm$^{-2}$~s$^{-1}$~arcsec$^{-2}$.  
The highest contour is set at 95\% the emission peak and the step between contours is constant for each panel.  
The spatial resolution of these maps is $\simeq1\farcs0$. 
}
\vspace*{-0.25cm}
\label{fig.MEGARA}
\end{center}
\end{figure*}

The velocity expansion along the line of sight of the [N~{\sc ii}]-bright inner regions of HuBi\,1 can be investigated using the 
MEGARA IFS  observations.
The [N~{\sc ii}] $\lambda$6584 emission line profile of the whole region 
shown in Figure~\ref{fig:spec1D} unveils broad wings indicative of a fast velocity component with a full-width at zero intensity (FWZI) $\approx500$ km~s$^{-1}$.
The blue (approaching) wing of this profile is noticeably brighter than the red (receding) one.

The spatial distribution of this fast component is uniquely revealed by MEGARA's tomographic capabilities.  
According to the velocity ranges defined in the [N~{\sc ii}] line profile in Figure~\ref{fig:spec1D}, nine velocity channels are shown in Figure~\ref{fig.MEGARA}: 
four 50 km~s$^{-1}$ in width channels mapping the blue component and another four the red component, and one 100 km~s$^{-1}$ in width channel mapping the emission at the systemic velocity. 
The blue component has a compact appearance, with the emission peaking in all velocity channels at an angular distance $\approx0\farcs3$ northeast (PA$\simeq$50$^\circ$) of the nebula center.  
On the other hand, the red component shows an arrowhead shape in the two channels closer to the systemic velocity, with its emission peaking at 
$\approx1\farcs0$ towards the southwest of the center of the nebula.
In the next two ``reddest'' channels, the emission looks more alike that of the blue component, with the emission peak close to the nebular center.  
The emission is extended, with a spatial extent that decreases along with the difference with the systemic velocity both for the blue and red components.

As for the emission at systemic velocities (0 km~s$^{-1} < v_{\rm hel} < +100$ km~s$^{-1}$), it can be described by a donut-shaped morphology, similar to the [N~{\sc ii}] image in Figure~\ref{fig:img}, as the emission from this velocity channel dominates the total emission (Fig.~\ref{fig:spec1D}).
The emission in this channel shows two peaks $\simeq1\farcs9$ apart along PA $\approx50^\circ$.

\begin{figure}
\begin{center}
\includegraphics[width=0.95\linewidth]{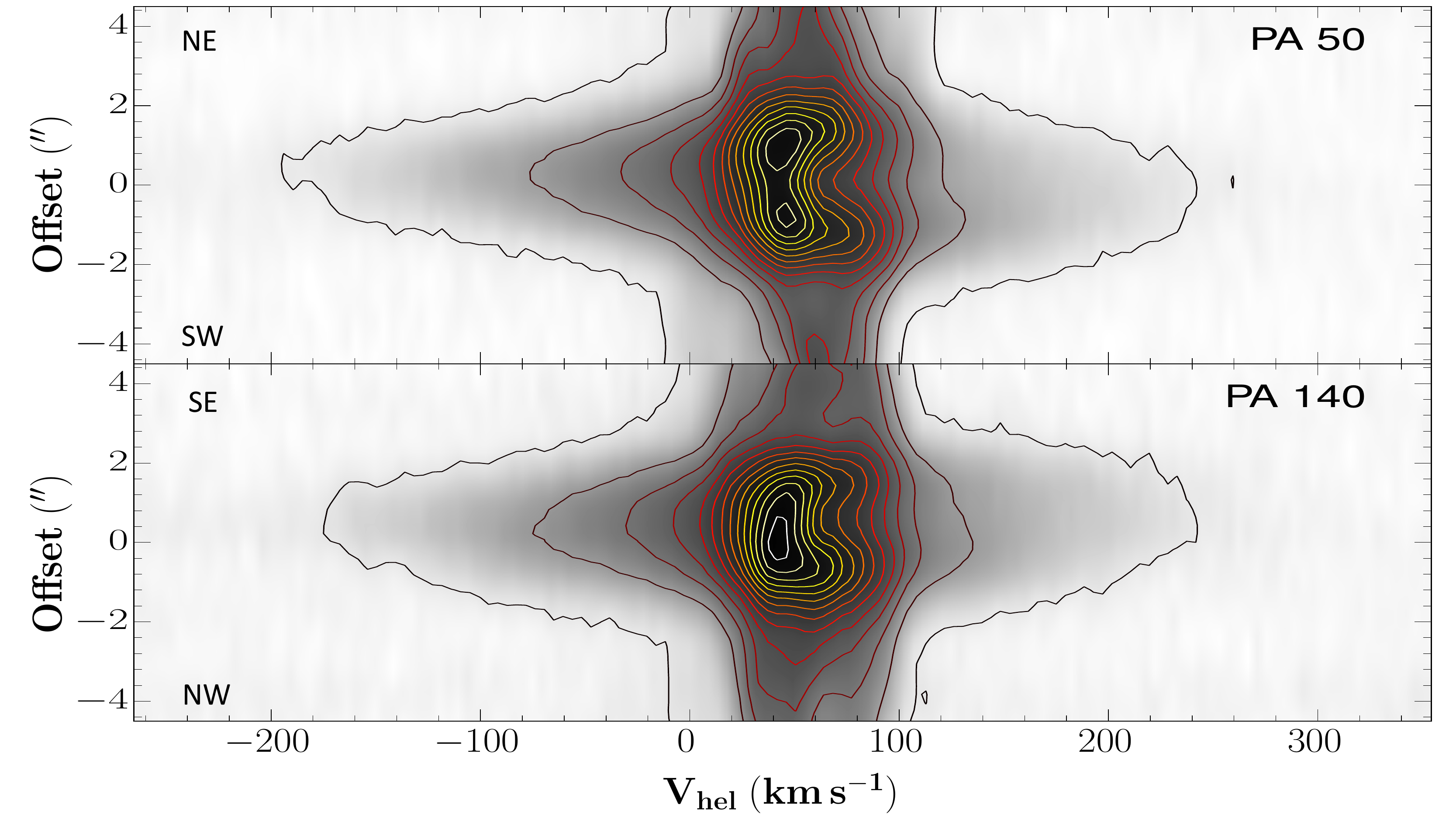}
\caption{
PV maps of the [N\,{\sc ii}] emission line extracted from MEGARA pseudo-slits at PA 50$^{\circ}$ and PA 140$^{\circ}$.
Contours of different colors are used to highlight both faint and bright emission.  
}
\vspace*{-0.45cm}
\label{fig.PV}
\end{center}
\end{figure}

The MEGARA IFS observations can be used to extract position-velocity (PV) maps along any PA of interest.  
In particular, the preferential direction along PA 50$^{\circ}$ and the orthogonal direction at PA 140$^{\circ}$ have been selected to extract PV maps from pseudo-slits in the MEGARA data cube (Fig.~\ref{fig.PV}).  
The broad wings from the fast velocity component are present in these PV maps, that confirm that the blue wing is brighter than the red one. 
In addition, these PV maps reveal a notable difference between the blue and red wings:  
the blue component of the outflow ($-200$ km~s$^{-1} \leq V_{\rm hel} \leq 0$ km~s$^{-1}$) has a Gaussian profile along the spatial direction, but the red component has diminished emission at offsets $\simeq 0^{\prime\prime}$ in the velocity range 100 km~s$^{-1} \leq V_{\rm hel} \leq 200$ km~s$^{-1}$.  
This is consistent with the spatial distribution of the emission shown in panels 1-4 and 6-9 in Figure~\ref{fig.MEGARA}.

\section{Discussion} 
\label{sec:discussion}

The new data presented above allow us to investigate key aspects of the innermost regions of HuBi\,1 to test the {\it born-again} scenario proposed by \citet{Guerrero2018}.  
One of their most basic predictions consisted in the presence of a fast ejecta that would shock-heat the material in the outer shell, the old PN.  
This is confirmed by the high-dispersion MEGARA IFS observations that detect material with radial expansion velocities up to 250 km~s$^{-1}$, but also by the expansion velocity on the plane of the sky $\simeq290$ km~s$^{-1}$ implied by the angular expansion rate.  
This velocity exceeds notably the shock-velocity range of 70--100 km~s$^{-1}$ required for the ionization of He$^{++}$ and O$^{++}$ according to MAPPINGS \citep{SD2017} models of the ionization structure of the inner shell.  
This would suggest that material from the old nebular shell is entrained by the new ejecta, thus reducing the shock-velocity.



The angular expansion rate $\simeq0\farcs0115$~yr$^{-1}$ derived from multi-epoch [N~{\sc ii}] images of the inner shell and its present 2\farcs3 radius imply an age $\approx200$ yr assuming a constant expansion velocity.  
This estimate is in excellent agreement with the 90--220 yr recombination time-scale of the outer shell derived by \citet{Guerrero2018}.

Finally, it is possible to assess the spatio-kinematic structure of the ejecta in HuBi\,1.  
Contrary to the kinematics of the equatorial disk and fast bipolar outflow detected in other {\it born-again} PNe \citep[e.g., A\,30,][]{Chu1997}, the expansion velocity along the line of sight of HuBi\,1 derived from the [N~{\sc ii}] FWZI, $\lesssim250$ km~s$^{-1}$, is quite similar to that derived on the plane of the sky, 
290$\times$($d$/5.3 kpc) km~s$^{-1}$. 
Furthermore, the tomography of a collimated outflow is expected to show a compact component at a single radial velocity for a knot or in a range of radial velocities for a filament with varying velocity. 
Instead, the tomography of the innermost regions of HuBi\,1 in Figure~\ref{fig.MEGARA} reveals that the emission is resolved, with its spatial extent decreasing as larger expansion velocities are mapped.  
This is exactly the expectation for an expanding shell, as the tomography maps the largest slices of an expanding shell on the plane of the sky and smaller slices at its poles.  
The similarity between the expansion velocity on the plane of the sky and along the line of sight and the small offset of the emission peaks in the most extreme velocity channels in Figure~\ref{fig.MEGARA} suggest that the inner region of HuBi\,1 is a shell with small ellipticity.

The spatio-kinematic properties of this shell reveal two interesting features.  
First, the receding component is fainter than the approaching component and its emission is diminished at positions close to that of the (now invisible) CSPN.  
This may be attributed to the high extinction at the CSPN location, absorbing the emission from the receding section of the shell. 
Second, there is a large brightness increase in the channel map at the systemic velocity.  
This may imply an over-density on the plane of the sky, i.e., an equatorial enhancement of material.  
The detailed distribution of material within this shell is thus uncertain.

The asymmetric ejecta in {\it born-again} PNe is not well understood, but it has been suggested a close binary interaction producing a nova-like eruption \citep[see][and references therein]{Wesson2018}. 
The recent discovery of a possible companion to the CSPN of A\,30 \citep{Jacoby2020} could support this hypothesis.  
In this sense, the shell-like distribution of the ejecta of HuBi\,1  and its longer evolution time, as compared to that of the CSPNe of the most recent {\it born-again} PNe A\,58 and Sakurai's object, seem to suggest that the VLTP event in HuBi\,1 was peculiar. 
This goes in line with the lower mass progenitor of HuBi1\,1 compared to that of other {\it born-again} PNe suggested by \citet{Guerrero2018}.

\section{SUMMARY} 
\label{sec:summary}

We have analyzed multi-epoch images and spatially-resolved IFS kinematic data of the innermost regions of HuBi\,1.  
The presence of a $\simeq300$ km~s$^{-1}$ high-velocity ejecta is confirmed, lending strong support to the {\it born-again} nature of HuBi\,1.  
The angular expansion rate derived from the comparison of multi-epoch images allows us to date this VLTP event $\simeq200$ yr ago.  
The spatial and spectral information simultaneously provided by the GTC MEGARA IFS observations have proven key to unravel the 3D spatio-kinematic structure of the [N~{\sc ii}]-bright innermost regions of HuBi\,1. 
Contrary to other {\it born-again} PNe, the ejecta in HuBi\,1 has a shell-like distribution.
This makes its inner shell the fastest among PNe, only superseded by PNe with extremely fast bipolar lobes. 



\acknowledgments
JSRG and VMAGG acknowledge support from the Programa de Becas 
posdoctorales of DGAPA UNAM (Mexico). JSRG, VMAGG, and JAT are funded by the 
UNAM DGAPA PAPIIT project IA100720 (Mexico). 
MAG acknowledges support of the Spanish Ministerio de Ciencia, Innovaci\'on 
y Universidades (MCIU) grant PGC2018-102184-B-I00.
MAG, SC and LFM acknowledge financial support from the State 
Agency for Research of the Spanish MCIU through the ``Center of Excellence Severo Ochoa'' 
award to the Instituto de Astrof\'\i sica de Andaluc\'\i a (SEV-2017-0709). 
ES and GRL acknowledge support from CONACyT. 
LFM is partially supported by MCIU grant AYA2017-84390-C2-1-R. 
JSRG and VMAGG thank INAOE for its hospitality during a 
training course on reduction and analysis of MEGARA data and the organizers  
of this course Y.\ D.\ Mayya and J.\ Zaragoza-Cardiel for sharing their expertise 
with us. The GTC Science Operations team is acknowledged for scheduling the 
GTC MEGARA observations under the stringent conditions demanded by this program.


\facilities{Nordic Optical Telescope (ALFOSC), Gran Telescopio de Canarias (MEGARA).}
\software{megararss2cube, IRAF}


\end{document}